\documentclass[journal]{IEEEtran}

\IEEEoverridecommandlockouts
\usepackage{amssymb}
\usepackage{nomencl}
\makenomenclature
\usepackage{amsmath,amssymb,amsfonts}
\usepackage{algorithmic}
\usepackage{graphicx}
\usepackage{textcomp}
\usepackage{xcolor}
\usepackage{svg}
\usepackage{lettrine}
\usepackage{comment}
\usepackage{scalerel}
\usepackage{tikz}
\usetikzlibrary{svg.path}

\usepackage{array}
\renewcommand*{\arraystretch}{1.0}
\usepackage[sorting=none]{biblatex} 
\def\BibTeX{{\rm B\kern-.05em{\sc i\kern-.025em b}\kern-.08em
    T\kern-.1667em\lower.7ex\hbox{E}\kern-.125emX}}
\usepackage{etoolbox}
\renewcommand\nomgroup[1]{%
  \item[\bfseries
  \ifstrequal{#1}{A}{Indices}{%
  \ifstrequal{#1}{B}{Parameters}{%
  \ifstrequal{#1}{C}{Sets}{%
  \ifstrequal{#1}{D}{Variables}{}}}}%
]}

\usepackage[hidelinks]{hyperref}
\usepackage[capitalise]{cleveref}

\begin{document}
\title{Strategic Planning of Carbon-Neutral Heating Demand Coverage Under Uncertainty in a Coupled Multi-Energy Grid\\
{\footnotesize }
\thanks{This paper has been developed as part of Integrated Network Planning (iNeP) of the research project Northern German Living Lab (\textit{Norddeutsches Reallabor}), which is supported by the German Federal Ministry of Economic Affairs and Climate Action (BMWK) under agreement no. 03EWR007O2.}
}

\makeatletter
\newcommand{\linebreakand}{%
  \end{@IEEEauthorhalign}
  \hfill\mbox{}\par
  \mbox{}\hfill\begin{@IEEEauthorhalign}
}
\author{\IEEEauthorblockN{Marwan Mostafa, Davood Babazadeh, and Christian Becker}\\
\IEEEauthorblockA{\textit{Institute of  Electrical Power and Energy Technology, Hamburg University of Technology,} Hamburg, Germany \\
marwan.mostafa@tuhh.de}
}

\maketitle

\begin{abstract}
Integrating the gas and district heating with the electrical grid in a multi-energy grid has been shown to provide flexibility and prevent bottlenecks in the operation of electrical distribution grids. This integration however assumes a top-down grid planning approach and a perfect knowledge of consumer behaviour. In reality, the consumer decides whether to adopt a heating technology based on costs and government regulation. This behavior is highly uncertain and depends on fluctuations in heating technology costs and energy prices. The uncertainty associated with consumer behavior increases the risk of investment in grid expansion. In response to this challenge, this paper proposes an approach with the consumer at the center of the planning method. Robust optimization is used to model the uncertainty in prices to reduce the risk of investment in grid expansion. The uncertainty in energy prices is modeled using interval uncertainty with a proportional deviation. This allows planners, operators and regulators to predict the adoption rate of certain heating technology in different geographical areas and prioritize the expansion of specific grids where they are required. By minimizing a cost function subject to robust constraints, the strategy ensures robustness against uncertainties in energy prices. This robust optimization approach is applied to Hamburg as a case study. 
The result of the optimization represents the consumer's decision. The impact of the consumer's decision on the electrical grid is analzed on different benchmark distribution grids. The study concludes that district heating expansion in high-density areas is a low-risk investment for carbon neutrality. In less dense areas, electrification supports decentralized heat pumps. Meanwhile, hydrogen gas grids are viable where electric expansion is impractical. Increased uncertainty leads to more conservative solutions. The results also show that putting the consumer instead of the planner at the center of the planning method results in more critical scenarios for grid expansion. This approach can be implemented promptly and practically by grid planners and is an important component of an integrated planning process for multi-energy grids.
\end{abstract}


\newcommand{\minus}{\scalebox{0.8}{$-$}}
\newcommand{\plus}{\scalebox{0.6}{$+$}}

\nomenclature[A,1]{\(i\)}{Cell index}
\nomenclature[A,2]{\(j\)}{Heating technology index}
\nomenclature[A]{\(de\)}{Decentralized heat pumps in households}
\nomenclature[A]{\(dg\)}{Decentralized hydrogen boilers in households}
\nomenclature[A]{\(ce\)}{District heating with centralized electric heat pumps}
\nomenclature[A]{\(cg\)}{District heating with centralized hydrogen boilers}

\nomenclature[B,1]{\(N\)}{Total number of cells}
\nomenclature[B,2]{\({H}\)}{Total heating energy demand per year in kWh/a}
\nomenclature[B,3]{\({P}\)}{Total peak heating load power in kW}
\nomenclature[B,4]{\({L}\)}{Total length of streets in m}
\nomenclature[B,5]{\({W}\)}{Binary flag for existing district heating grid}
\nomenclature[B,6]{\({C^{l}}\)}{Cost of grid expansion in €/m or €/kW}
\nomenclature[B,7]{\({C^{p}}\)}{Cost of heating power generator in €/kW}
\nomenclature[B,8]{\({\hat{C}^{f}}\)}{Predicted price of energy carrier}
\nomenclature[B,90]{\({\tilde{C}^{f}}\)}{Uncertain price of energy carrier}
\nomenclature[B,91]{\({\Delta^{f}}\)}{Maximum deviation in price of energy carrier}
\nomenclature[B,92]{\({\eta}\)}{Efficiency of heating power generator}
\nomenclature[B,93]{\({\mathrm{COP}}\)}{Coefficient of performance of electric heat pumps}
\nomenclature[B]{\({\sigma}\)}{Scale factor for grid expansion}
\nomenclature[B]{\({\bar{\Gamma}}\)}{Upper limit on grid expansion}

\nomenclature[C]{\({\mathcal{J}}\)}{Set of heating technologies, \({\mathcal{J} = \{ce,cg,de,dg\}}\)}
\nomenclature[C]{\({\mathcal{U}}\)}{Uncertainty set of energy carrier price}

\nomenclature[D]{\({X}\)}{Binary decision variable}

\printnomenclature
\thispagestyle{empty}
\pagestyle{empty}

\section{Introduction}\label{intro}

Achieving carbon neutrality in energy systems requires innovative approaches in the integrated planning of distribution grids, including electricity, gas, and district heating. A main challenge in achieving this goal is the prioritization of grid expansion in different geographical areas. Conventional planning methods typically use a top-down approach in forecasting future grid needs. These conventional methods pay little attention to consumer behavior and assume perfect knowledge in their adaptation of new technologies. However, the consumer's decision to adopt a new technology is motivated by financial interest or government regulation.

The city of Hamburg has set the target of carbon neutrality by 2045. To meet this target, an integrated planning process was proposed in \cite{mostafa_integrated_2022-1}. In this process, the optimal grid expansion roadmap guaranteeing the security of a carbon-neutral energy supply is to be found. The first step in this process is to forecast future heating energy demand as heating is the dominant end-use of the transported energy through the grid. The heating energy demand and load distribution are aggregated in cells which offer a comprehensive view of energy systems, incorporating energy demand and various socio-economic factors. This method provides detailed regional demand forecasts and structures \cite{vorwerk_multifaceted_2023}. In this study, a 1 km\textsuperscript{2} grid represents these cells.

It was found in \cite{heise_coupled_2023} that the provision of heat pumps at the low-voltage level cannot be implemented without significant grid expansion. In addition to heat pumps, the integration of additional loads, notably electric vehicles, into the grid due to increased electrification in all sectors intensifies the existing challenges. It was demonstrated that a centralized solution through local or district heating can alleviate the time-critical expansion in the electrical grid. Additionally, centralized heat supply offers the opportunity to integrate short-term and seasonal storage. These, in turn, allow for a notably enhanced flexibility in grid operation, presenting a significant advantage. At the coupling elements between the heating and electrical grids, the provided flexibility can be used to support the operation of the electrical grid during periods of high utilization. 

The next important step is to identify potential areas for centralized heat supply and at the same time reduce the immense expansion needs of the electrical grid. This however must be done considering the consumers decision and government regulation. In 2024 the government passed the new buildings energy act \cite{german_federal_government_new_nodate}. In this paper, four heating technologies associated with four different grid expansion options are studied and the decision to invest in regionally expanding one of the four grid options is taken. The grid associated with the lowest total economic cost must be chosen. This decision is made under uncertain energy prices and during a dynamic and crucial transition phase in the grid as investing in expanding the wrong grid might hinder achieving the carbon neutrality goal, increasing the investment risk considerably.

Scenario-based stochastic optimization methods are widely used to handle uncertainties in power generation and energy prices in the planning of electrical distribution grids \cite{hakami_review_2022}. According to \cite{amjady_adaptive_2018,resener_electric_2020}, RO can be more efficient than these approaches in solving planning problems. This is because unlike in stochastic optimization where the solution space depends on the number of uncertain parameters and the number of scenarios, in RO the size of the solution space depends on the number of uncertain parameters only. It is shown in \cite{mohammadi-ivatloo_robust_2019} that a two-stage RO approach for the short-term planning of an electrical distribution grid is an efficient tool for solving complex planning problems. Most applications of RO have focused on the optimization of energy systems \cite{kabiri-renani_robust_2023,roald_power_2023}, sizing of batteries and renewable generation \cite{aghamohamadi_two-stage_2021,li_two-stage_2021}.
Moreover, the implementation of RO in the planning is limited to uncertainties in the dispatching and renewable generation and considers only the electrical distribution grid. 

This work's aim is to expand these methods to a multi-energy distribution grid and increase the number of uncertain parameters. The goal of the RO problem is to find the optimal grid expansion decision that minimizes a cost function subject to various constraints while ensuring robustness against uncertainty. This RO technique provides a solution that is optimal for the worst-case scenario and feasible for all realizations of uncertainties. This approach provides multi-energy grid planning steps that are robust against the uncertainty in the parameters. It guarantees a satisfactory performance under all possible scenarios ensuring that the subsequent planning steps in the multi-energy grid are not significantly degraded even under unfavorable conditions.

This paper presents a new planning method that puts the consumer at the center. The consumer's financial interest is represented as a cost minimization optimization problem. Robust optimization (RO) is used to address the uncertainties of future energy prices. The consequences of the consumer's decision on the grid is analyzed using power flow calculations.

In Section \ref{deterministicmodel}, the problem is formulated as a deterministic optimization problem. The proposed RO problem is introduced in Section \ref{robustmodel}. Section \ref{resultsanddiscussion} presents the numerical result. In Section \ref{conclusion}, conclusions are drawn and an outlook for the expansion of this model is discussed.

\section{Deterministic Model} \label{deterministicmodel}
\begin{figure*}[htbp]
\centerline{\includegraphics[width=\textwidth,height=\textheight,keepaspectratio]{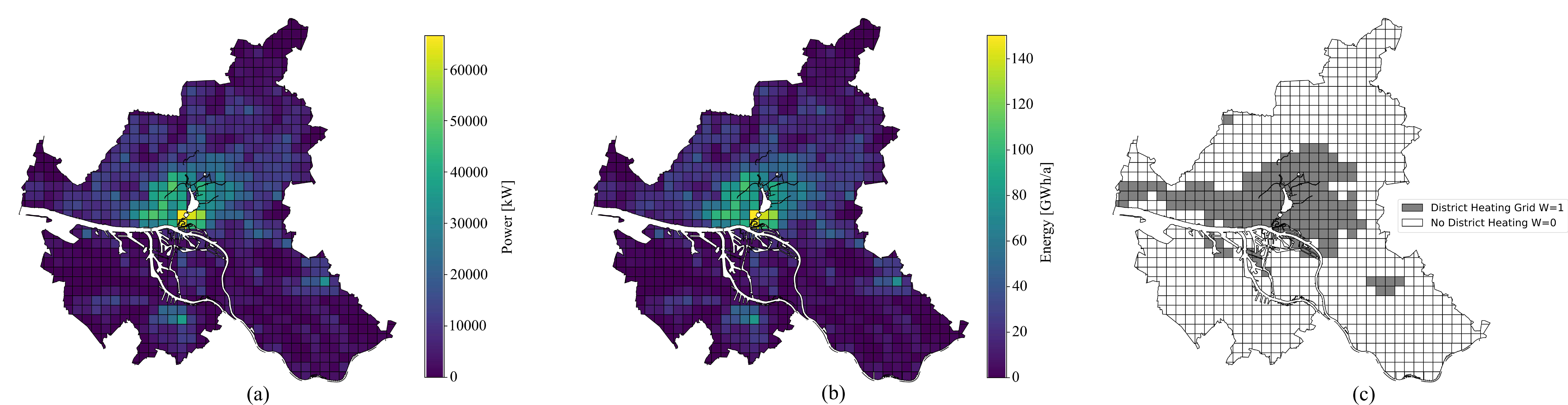}}
\caption{Aggregated data for heating in Hamburg in the year 2045. (a) Peak load power based on standard load profiles in kW, (b) Heating energy demand per year kWh/a, (c) Coverage of the current district heating grid.}
\label{widefig}
\end{figure*}
A deterministic cost optimization model to choose the cost-optimal grid expansion in each cell is presented. Fig. \ref{widefig} shows the input data to this model represented in (a) aggregated data for peak heating load power based on standard load profiles for Germany \cite{meier_repraesentative_1999}, (b) total heating energy consumption per year and (c) current district heating grid coverage.
\subsection{Objective Function}
The objective function (OF) of the proposed model formalized in \eqref{eq1:main} minimizes the total economic cost associated with implementing one of four technologies:  decentralized household heat pumps, decentralized household hydrogen boilers, district heating with centralized electric heat pumps and district heating with hydrogen boilers. 
\begin{subequations}\label{eq1:main}
\begin{equation}
min\ \sum_{i=0}^{N}\sum_{j \in \mathcal{J}}  X_{i,j}\cdot \hat{C}_{i,j}  \label{eq1:a}
\end{equation}
The cost represented in \eqref{eq1:b}  consists of three components: the cost of energy, the cost of the heating power generator and the cost of grid expansion.
The energy cost depends on the total yearly energy demand in kWh/a and the efficiency of the heating technology (COP for heat pumps). The cost of power generators depends on the load power and the cost of heat generators per kW. The cost of grid expansion for gas and district heating depends on the length of the streets in a cell modified by a factor $\sigma$ derived from empirical analysis of the length of the existing grids compared to the length of the streets. This reflects the fact that gas and district heating grids are laid along the streets. Electrical distribution grid expansion however depends on the grid capacity in kW instead of the length of the streets.
\begin{multline}    
\hat{C}_{i,j} = \underbrace{\frac{\hat{C}^{f}_{j}\cdot H_{i}}{\eta_{j}}}_{\text{Energy cost}} + \underbrace{{C}^{p}_{j}\cdot P_{i}}_{\text{Generator cost}}\\ + \underbrace{{C}^{l}_{j}\cdot \sigma_{j}\cdot \{
\begin{array}{ll}
L{i} \quad &  j \neq de \\
P_{i} \quad & j = de \
\end{array}
\}}_{\text{Grid expansion cost}}  \quad;\forall i, \forall j \in \mathcal{J} \label{eq1:b}
\end{multline}
\subsection{Constraints}
In \eqref{eq1:c} and \eqref{eq1:d} the binary integer decisions are constrained such that only one grid is selected for each cell.
\begin{equation}
X_{i,j} \in \{0,1\} \qquad;\forall i, \forall j \in \mathcal{J}\label{eq1:c}
\end{equation}
\begin{equation}
\sum_{j \in \mathcal{J}} X_{i,j} = 1 \qquad;\forall i \label{eq1:d}
\end{equation}
To consider the existing district heating grid in our solution, a constraint is added to guarantee that one of the two centralized heating technologies will be used in a cell if it already has a district heating grid. This if-statement can be expressed as a linear program using the Big M method in \eqref{eq1:e} and \eqref{eq1:f}, where $M$ is a large number.
\begin{equation}
P_{i} \geq Q  - M \left(1 - X_{i,j} + W_{i} \right)\quad;\forall i , \forall j \in \{ce,cg\} \label{eq1:e}
\end{equation}
\begin{equation}
X_{i,j}  \leq  1 - W_{i} \quad;\forall i, \forall j \in \{de,dg\}  \label{eq1:f}
\end{equation}
Lastly, the investment in electric grid expansion is limited in \eqref{eq1:g} to a maximum value $\Bar{\Gamma}_{de}$ to reflect practical expansion limits within the given timeframe and other electrification needs in the city.
\begin{equation}
\sum_{i=0}^{N}X_{i,de} \cdot P_{i} \leq \Bar{\Gamma}_{de}  \label{eq1:g}
\end{equation}
\end{subequations}

\section{Robust Model} \label{robustmodel}
As mentioned in section \ref{intro}, multiple factors can introduce uncertainty in our problem. However, this work focuses on modeling the uncertainty of energy prices only. This is because energy costs constitute the major share of the total economic cost and historically, energy prices have fluctuated greatly due to technological advancements and geopolitical events.

\subsection{Uncertainty Set Realization}
The proportional interval uncertainty set\cite{bertsimas_price_2004} presented in \eqref{eq2} represents the uncertainty associated with the fluctuation in energy carrier prices. The uncertain values can deviate from the predicted or nominal values by a deviation factor $\Delta$ which is given as a percentage of the predicted value.
\begin{equation}
\mathcal{U}_{j}= \{ \Tilde{C}^{f}_{j} \in \mathbb{R}_{+}:\Big|\frac{\Tilde{C}^{f}_{j}- \hat{C}^{f}_{j}}{\hat{C}^{f}_{j}}\Big| \leq \Delta^{f}_{j}\} \quad; \forall j \in \mathcal{J} \label{eq2}
\end{equation}
For the purpose of this research, only two energy carriers, electricity and hydrogen, are taken into account, resulting in
$\Delta^{f}_{de}=\Delta^{f}_{ce}=\Delta^{f}_{electricity}$ and $\Delta^{f}_{dg}=\Delta^{f}_{cg}=\Delta^{f}_{hydrogen}$. 
.

\subsection{Robust Formulation} 
Expanding the deterministic model in \eqref{eq1:main} with the uncertainty set in \eqref{eq2} results in the robust formulation in \eqref{eq3:main}. The resulting formulation is a max-min problem. The outer maximization problem represents the realization of the uncertainty and achieves the worst-case scenario for energy carrier prices. The inner minimization problem has the objective of choosing the decision variables that achieve the minimum cost under this worst-case scenario.
\begin{subequations}\label{eq3:main}
\begin{equation}
 \min_{X}\ \max_{\tilde{C}^{f}} \sum_{i=0}^{N}\sum_{j \in \mathcal{J}}  X_{i,j}\cdot \tilde{C}_{i,j}  \label{eq3:a}
\end{equation}
subject to:\\
\begin{equation}
\tilde{C}^{f}_{j} \in \mathcal{U}_{j} \qquad; \forall j \in \mathcal{J}   \label{eq3:c}
\end{equation}
\begin{equation}
\eqref{eq1:c}\mbox{--}\eqref{eq1:g}, \eqref{eq2}  \label{eq3:e}
\end{equation}
\end{subequations}

\section{Results and Discussion} \label{resultsanddiscussion}
The study in this paper is conducted for a future scenario of the year 2045. In each cell, buildings are classified into different categories such as single-family houses, apartment buildings, commercial buildings, mixed-use buildings and public buildings. Based on the size of the buildings, state of renovation and population density in a cell, the heat energy and power demand are forecasted and aggregated in each cell \cite{mostafa_integrated_2022-1, vorwerk_multifaceted_2023, dochev_assigning_2020}. The existing power generation capacities are assumed not to be carbon-neutral and are therefore not considered.
\subsection{Data Set}
Alongside the previously mentioned aggregated data in fig. \ref{widefig}, the problem parameters are complemented with the data in table \ref{tab1}. The cost of heating power generators is divided over their lifetime of 20 years. These costs are estimated based on internal studies done by industry experts from our project partners. The maximum deviation in the price of electricity is set to 50 \%. In contrast, the maximum deviation in the price of green hydrogen is set to 200 \% to reflect the greater uncertainty in the predicted prices of green hydrogen, due to the lack of a green hydrogen market today and the unavailability of its mass production. \cite{wintzek_planning_2021}

\renewcommand{\arraystretch}{1.3} 
\begin{table}[htbp]
\centering
\caption{Summary of data set parameters}
\begin{tabular}{|c|c||c|c|}
\hline
\textbf{Parameter}&\textbf{Value}&\textbf{Parameter}&\textbf{Value}\\
\hline
$\mathrm{COP}_{ce}$ &3.8&$\mathrm{COP}_{de}$ &3.0\\
\hline
$\eta_{cg,dg}$&88\%&$\bar{\Gamma}_{de}$&1.2 GW\\
\hline
$C_{dg}^{l}$&610 €/m&$C_{ce,cg}^{l}$&2160 €/m\\
\hline
$C_{de}^{l}$&320 €/kW&$\sigma_{ce,cg}$&1.1\\
\hline
$\sigma_{dg}$&1.2&$\sigma_{de}$&6.0\\
\hline
$C_{ce}^{p}$&798 €/kW&$C_{de}^{p}$&1270 €/kW\\
\hline
$C_{cg}^{p}$&79 €/kW&$C_{dg}^{p}$&235 €/kW\\
\hline
$\hat{C}_{ce}^{f}$&15.2 ct/kWh&$\hat{C}_{de}^{f}$&23.7 ct/kWh\\
\hline
$\hat{C}_{cg}^{f}$&10.0 ct/kWh&$\hat{C}_{dg}^{f}$&10.7 ct/kWh\\
\hline
$\Delta_{electricity}^{f}$&50\%&$\Delta_{hydrogen}^{f}$&200\%\\
\hline
\end{tabular}
\label{tab1}
\end{table}

\subsection{Numerical Results}
The robust mixed-integer linear programming (MILP) formulation was solved using the MOSEK solver on Apple M1 16 GB RAM in Python.
The results shown in fig. \ref{resultsmap} show that the bulk of heat demand should be covered by district heating using centralized heat pumps. In lower-density areas, the electrical distribution grid should be expanded to cover the heating demand using decentralized household heat pumps. In areas with very low load density, on the other hand, the hydrogen gas grid should be developed as heating using decentralized household gas boilers is the most cost-effective solution. District heating using centralized hydrogen boilers is not chosen in any cell.

\begin{figure}[b]
\centerline{\includegraphics[width=0.5\textwidth,height=\textheight,keepaspectratio]{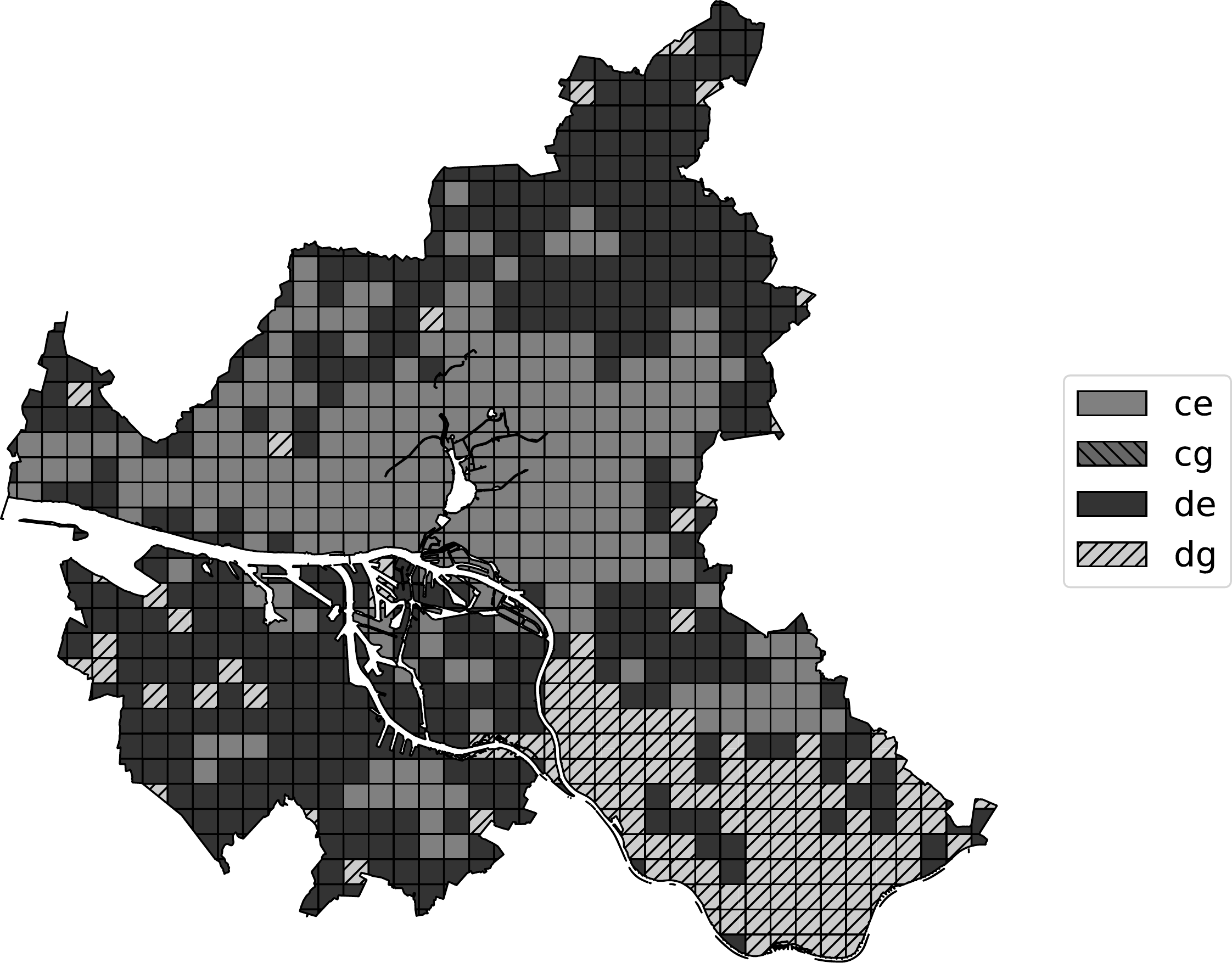}}
\caption{Solution of the robust optimization problems showing the distribution of heating technologies.}
\label{resultsmap}
\end{figure}

\subsection{Post-event Analysis}
To analyze the effect of the size of the uncertainty set on the results, a post-event analysis was conducted. In the analysis, the optimization problem was solved for increasing values of maximum deviation in the hydrogen price from 0\% to 200\% and the maximum deviation in the price of electricity was kept constant at 50\%. The result of the analysis shown in fig. \ref{resultstechnology} shows that at 0\% deviation the dominant chosen technologies are by far those depending on hydrogen as an energy carrier but as soon as uncertainty is introduced the decentralized heat pumps dominate the low-density areas and the maximum electric grid expansion of 1.2 GW is utilized. From a deviation value of 70\% and higher, the district heating solution with electric heat pumps becomes dominant in covering most of the demand. This clearly shows that investing in the district heating grid is a low-risk decision since it is cost-effective in most of the higher-load density areas in the city regardless of the deviation value and the energy carrier used.

The impact of the size of the uncertainty set on the investment cost in the energy infrastructure, which includes the grid expansion and generator costs is shown in fig. \ref{resultsmoney}. We can see an inherent characteristic of RO. As the size of the uncertainty set increases, the solution becomes more conservative and therefore the investment costs needed increase. The analysis indicates that when the deviation grows to 10\% the costs increase by 23\% compared to the initial value at 0\% deviation. Beyond 70\% deviation, the costs increase by 38\% over the initial value.

\begin{figure}[htbp]
\centerline{\includegraphics[width=0.5\textwidth,height=\textheight,keepaspectratio]{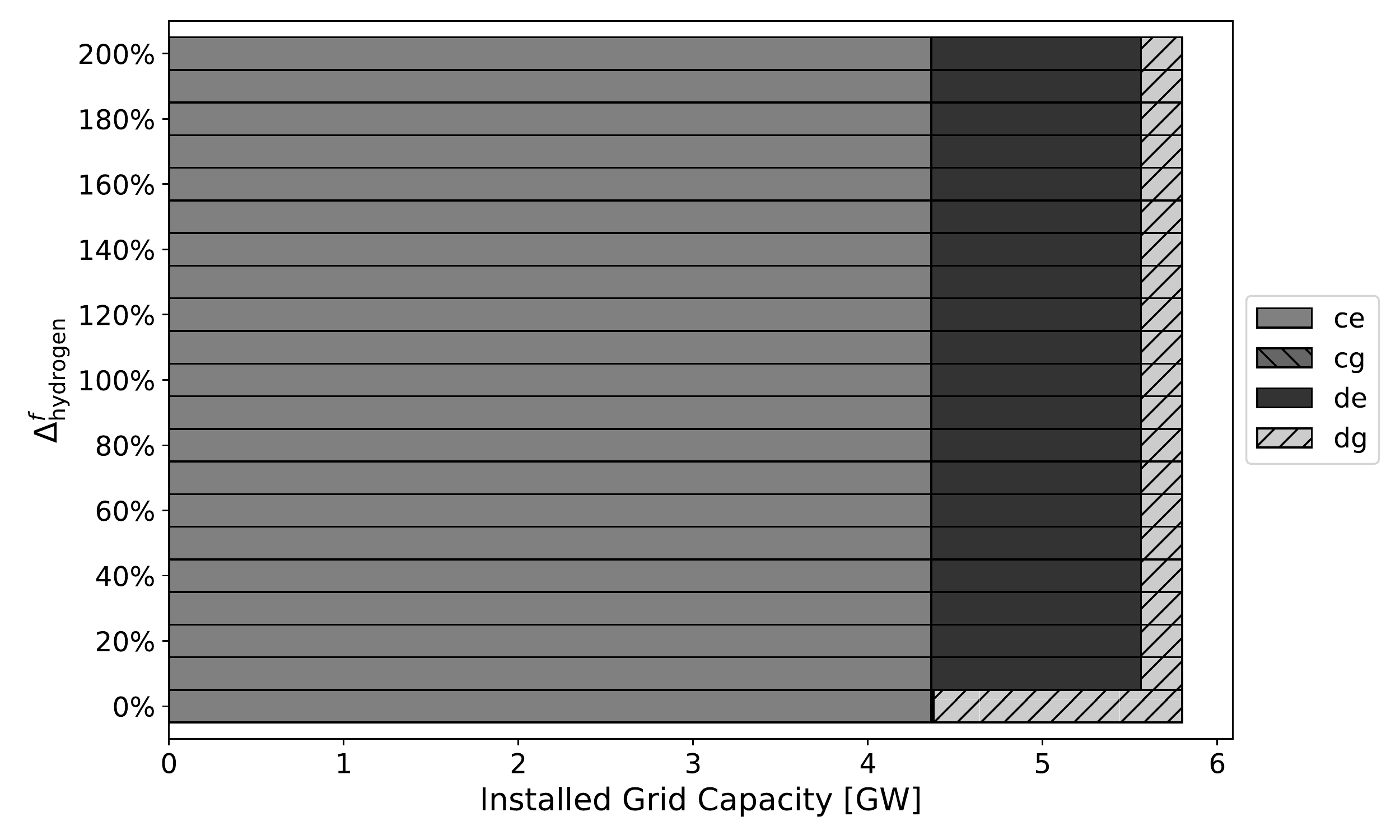}}
\caption{Effect of the size of the uncertainty set due to increasing deviation from the predicted hydrogen prices on the installed grid capacity.}
\label{resultstechnology}
\end{figure}

\begin{figure}[htbp]
\centerline{\includegraphics[width=0.5\textwidth,height=\textheight,keepaspectratio]{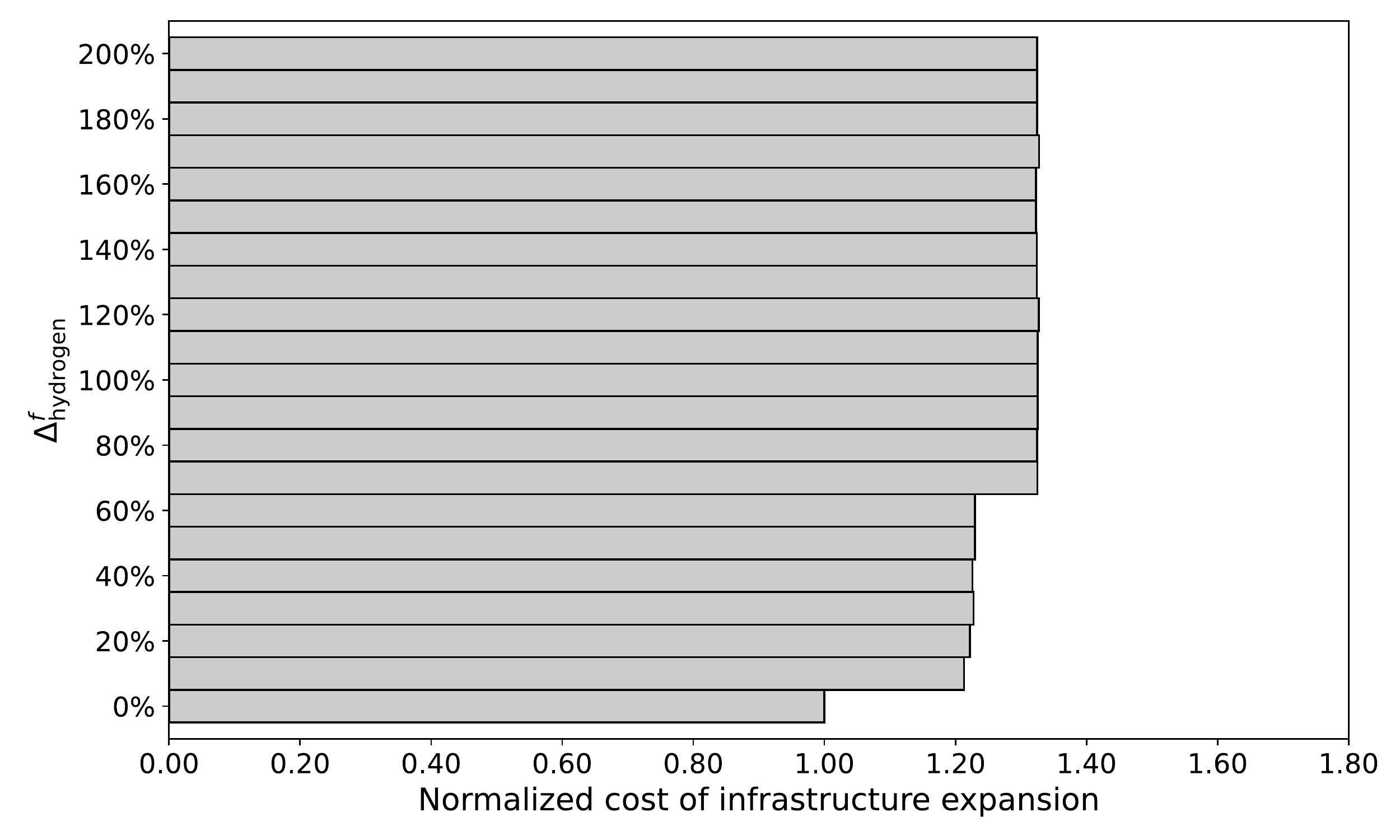}}
\caption{Effect of the size of the uncertainty set on the investment cost in infrastructure.}
\label{resultsmoney}
\end{figure}
\section{Conclusion and Outlook} \label{conclusion}
This paper proposed a novel robust optimization approach to model the uncertainty of energy prices in grid expansion planning to aid decision-making in a new planning process for integrated multi-energy grids. The uncertainty in energy prices is modeled using interval uncertainty sets with a proportional deviation resulting in a box-constrained uncertainty. The new method provides significant benefits for strategic, integrated grid planning. Primarily it helps grid planners and regulating authorities make strategic decisions to prioritize the expansion of the individual grids in specific areas of the city with reduced investment risk despite the uncertainty. This approach can additionally be implemented directly and pragmatically by grid planners in their current planning processes. Furthermore, it is a main component of an innovative holistic integrated planning process for multi-energy grids.

According to the results expanding the district heating grid is a low-risk investment in areas with high load density. Conversely, in areas where load density is below 10 MW/km\textsuperscript{2}, the key to carbon neutrality in heating lies in the expansion of the electric grid, which will facilitate the adoption of decentralized household heat pumps. In areas where load density is low and where prioritizing electric grid expansion is not feasible or limited, the development of a hydrogen gas grid emerges as a strategic move. The results also show that the solution became more conservative as the uncertainty set became larger.  

The proposed approach shows potential for expansion by modeling more uncertain parameters and heating technologies as it has a short runtime despite the large amount of data. The next steps are to expand this method to model the more uncertain parameters as well as implement more detailed modeling of the uncertainties using distributionally robust and chance-constrained optimization methods. 

\section*{Acknowledgment}
The authors of this paper thank their industrial project partners Hamburger Energiewerke GmbH, Gasnetz Hamburg GmbH and Stromnetz Hamburg GmbH for their technical support and insight into the processes of grid operation and planning.

\printbibliography
\end{document}